\documentstyle[11pt,aaspp4]{article}

\def\etal{{\it et~al.}}
\def\bec{\begin{center}}
\def\enc{\end{center}}

\def\Mesz{M\'esz\'aros~}
\def\Pacz{Paczy\'nski~}
\def\nsns{NS-NS~}
\def\bhns{BH-NS~}
\def\simg{\mathrel{\hbox{\rlap{\lower.55ex \hbox {$\sim$}}
                   \kern-.3em \raise.4ex \hbox{$>$}}}}
\def\siml{\mathrel{\hbox{\rlap{\lower.55ex \hbox {$\sim$}}
                   \kern-.3em \raise.4ex \hbox{$<$}}}}
\def\Omj{\Omega_j}
\def\msun{M_\odot}
\def\Mbh{M_{bh}}
\def\eps{\epsilon}
\def\beq{\begin{equation}}
\def\enq{\end{equation}}
\begin{document}
\title{Afterglows from Gamma-ray Bursts\footnote{Invited Talk at the Symposium {\it Highlights in X-ray Astronomy} in honor of Joachim Truemper, Garching, Germany, 17-19 June 1998}  }

\author{P. \Mesz }  
\noindent
{Dpt. of Astronomy \& Astrophysics, 
Pennsylvania State University, University Park, PA 16802}

\begin{abstract}

The successful discovery of X-ray, optical and radio afterglows of gamma-ray
bursts has significantly helped our 
understanding of these sources, and made possible the identification of host 
galaxies at cosmological distances.  The energy release inferred in these 
outbursts rivals that of supernovae, while its photon energy output may 
considerably exceed it.  Current models envisage this to be the outcome of a 
cataclysmic stellar event leading to a relativistically expanding fireball, 
in which particles are accelerated at shocks and produce nonthermal radiation.
The substantial agreement between observations and the theoretical
predictions of the fireball shock model provide confirmation of the basic
aspects of this scenario. The continued observations show a diversity
of behavior, providing valuable constraints for more detailed models.
Crucial questions being now addressed are the beaming at different
energies and its implications for the energetics, the time structure
of the afterglow, its dependence on the central engine or progenitor system
behavior, and the role of the environment on the evolution of the afterglow.

\end{abstract}

\section{Introduction}

The first discovery of a GRB afterglow at X-ray wavelengths was made
with the BeppoSAX satellite (\cite{cos97}) in February 1997, followed 
by a long-term detection and follow-up at optical wavelengths (\cite{jvp97}).
This represents a major and long-awaited breakthrough in the investigation 
of these mysterious sources, and emphasizes the importance of high resolution 
X-ray imaging techniques, which were succesfully used for other significant
astronomical discoveries by Joachim Tr\"umper and his colleagues with Rosat.
The subsequent detection of other GRB afterglows followed in rapid succession, 
extending in some cases to radio and microwave wavelengths as well. This has 
made it possible to follow some of these sources over time scales of many 
months, making the identification of counterparts and host galaxies possible. 
The study of afterglows has provided strong confirmation for the generic 
fireball shock model of GRB, in which the $\gamma$-ray emission arises at 
radii of $10^{13}-10^{16}$ cm (\cite{rm92,mr93,rm94,px94,ka94,sapi95}). 
In particular, this model led to the prediction of the quantitative nature of 
the signatures of afterglows (\cite{mr97a}), in advance of the observational 
discoveries 
and in substantial agreement with these (\cite{vie97a,tav97,wax97,rei97,wrm97}).
The first measurement of a redshift (\cite{metz97}) in GRB 970508 provided 
confirmation of the hypothesis that bursts were at cosmological distances.
More recently, significant interest was aroused by the report (\cite{kul98}) 
of a spectroscopic redshift measurement of $z=3.4$ for the afterglow of the 
burst GRB 971214, whose fluence would then correspond to a $\gamma$-ray energy 
of $10^{53.5}(\Omega_\gamma /4\pi)$ ergs, where $\Delta \Omega_\gamma$ is the 
solid angle into which the gamma-rays are beamed.  
Such energies were discussed (\cite{mr97b}) in the context of compact 
mergers, such as neutron star-neutron star (\nsns) or black hole-neutron star 
(\bhns) mergers, which can power a relativistic fireball resulting in the
observed radiation. In some of the detected afterglows there is evidence
for a relatively dense gaseous environments, as
suggested, e.g. by evidence for dust (\cite{rei98}) in GRB970508,
the absence of an optical afterglow and presence of strong soft X-ray
absorption (\cite{gro97,mur97}) in GRB 970828, the lack an an optical 
afterglow in the (radio-detected) afterglow (\cite{tay97}) of GRB980329, and
spectral fits to the low energy portion of the X-ray afterglow of several
bursts (\cite{ow98}). The latter observations may be suggestive of ``hypernova" 
models (\cite{pac98,fw98}), involving the collapse of a massive star or its 
merger with a compact companion, although it is probably too early to draw
definite conclusions.  While it is at present unclear which, if any, of these 
progenitors 
is responsible for GRB, or whether perhaps different progenitors
represent different subclasses of GRB, there is general agreement that they
all would be expected to lead to the generic fireball shock scenario
mentioned above. Much of the current effort centers around trying to
identify such progenitors more specifically, and trying to determine what
effect, if any, they have on the observable fireball and afterglow 
characteristics.

\section{The Fireball Shock Scenario}

The isotropy of the angular distribution of GRB suggested already early on
that GRB were outside our own galaxy (\cite{fm95}), and the lack of structure
associated with nearby galaxies indicated that their distance would be such
that non-Euclidean and evolutionary effects may be important. This
was also indicated by studies of the counts of the number of bursts per 
unit fluence or peak flux (\cite{fm95,hor96}),
as did evidence for time dilation in the bursts (\cite{nor95}). Fits to
cosmological counts including curvature, luminosity function and evolution 
effects indicated that the latter could be important (\cite{rm97}). In
particular, the evolution may be related to the star formation rate as
a function of cosmological age (\cite{tot97,wij98,kru98}). The cosmological
distances of some bursts have, since then (\cite{metz97,kul98,djo98}) been
directly measured. This, of course, poses significant constraints on the
type of sources that may be able to produce this phenomenon. The typical 
numbers characterizing the luminosity, duration, total energy per burst and 
event rate are
\beq
L \sim 10^{52} (\Omega/4\pi) ~\hbox{erg~s$^{-1}$}~,
\enq
\beq
t_w \sim 10 ~\hbox{s}~,
\enq
\beq
E \sim 10^{53}(\Omega/4\pi)~\hbox{erg}~,
\enq
\beq
{\cal R} \sim 10^{-6} (\Omega/4\pi)^{-1}~\hbox{galaxy$^{-1}$yr$^{-1}$}~,
\enq
where $\Omega$ is the solid angle into which the energy is channeled, and the
durations $t_w$ have a large spread $10^{-3}\siml t_w \siml 10^3$ s. The
luminosities and total energies are also likely to be affected by a significant
spread, whose magnitude is harder to estimate. The beaming solid angle $\Omega_j$
is also, thus far, very poorly constrained.

An early suggestion for a GRB energy source at cosmological distances was 
the merger of \nsns binaries (e.g. \cite{pac86,goo86}), whose estimated numbers 
and merger rates are close to the observed GRB detection rate 
(\cite{nar91,phi91}). A related possibility are \bhns binary mergers
(\cite{nar91,pac91,mr92b}), which have only slightly different event rates.
An alternative suggestion (\cite{woo93}) was 
a ``failed Supernova Ib", resulting from the collapse of fast rotating 
massive star which fails to produce a core collapse SN. 
Both a compact merger, and a fast rotating stellar collapse were known,
from numerical simulations, to lead to a fast rotating torus around a
central, high density object which eventually would develop into a black hole.
The binding energy liberated in both of these events is of order $10^{54}$ 
ergs, a good fraction of which will be carried away in a short pulse
of $\nu\bar\nu$ and gravitational waves. If a small fraction of this 
emerges as electromagnetic
energy, some of which is deposited in a region with sufficiently low baryon
density, a relativistic fireball would form, whose radiation spectrum in
the observer frame will peak in the MeV range (\cite{pac86,goo86,sp90}). 
One possible channel for converting some of this energy into electromagnetic 
form is given by the $\nu\bar\nu \to e^+e^-$ process (\cite{eic89}). A low
baryon load condition (required to make the fireball highly relativistic) 
can occur naturally in a binary merger since a centrifugal barrier develops 
along the orbital symmetry axis, which is relatively free of baryons and 
provides an escape route for the $e^\pm,\gamma$ fireball (\cite{mr92}).  
This would also imply a collimation of the relativistic fireball, which 
would enhance the apparent flux (relative to what it would be if it were 
isotropic) by factors which conservatively might be of order 10-100.  
The same processes were estimated to be able to produce a collimated 
relativistic fireball in the failed SN Ib model (\cite{woo93}).
Another suggestion for powering a fireball was that this might result from
magnetic flaring activity (\cite{nar92}) in the disrupted torus produced
around the merged binary. 

Irrespective of the details of the progenitor, the resulting fireball
is expected to be initially highly  optically thick. From causality 
considerations the initial dimensions must be of order $c t_{var} \siml 10^7$ 
cm, where $t_{var}$ is the variability timescale,
and the luminosities must be much higher than a solar Eddington limit.
Since most of the spectral energy is observed above 0.5 MeV, the optical
depth against $\gamma\gamma \to e^\pm$ is large, and an $e^\pm,\gamma$
fireball is expected. Due to the highly super-Eddington luminosity, this
fireball must expand. Since in many bursts one observes a large fraction of 
the total energy at photon energies $\eps_\gamma \simg 1 GeV$, somehow the 
flow must be able to avoid degrading these photons ( since $\gamma\gamma 
\to e^\pm$ would lead, in a stationary or slowly expanding flow, to photons 
just below 0.511 MeV [\cite{hb94}]).
In order to avoid this, it seems inescapable that the flow must be expanding
with a very high Lorentz factor, since in this case the relative angle at which
the photons collide is less than $\Gamma^{-1}$ and the threshold for the
pair production is effectively diminished. Thus, photons with energy
\beq
\eps_{\gamma,\rm MeV} \siml 10^4 \eps_{t, \rm MeV}^{-1} \Gamma_2^2 ~.
\enq
are able to escape, where $\Gamma_2$ is the bulk Lorentz factor in units
of $10^2$, and $\eps_t$ is the energy of the target photons (\cite{m95}). Thus, 
simply from observations and general physical considerations, a relativistically
expanding fireball is expected. However, the observed $\gamma$-ray spectrum 
is generally a broken power law, i.e., highly nonthermal. The 
optically thick $e^\pm \gamma$ fireball cannot, by itself, produce such a 
spectrum (it would tend rather to produce a modified blackbody). In addition,
the expansion would lead to a conversion of internal energy into kinetic
energy of expansion, so even after the fireball becomes optically thin,
it would be highly inefficient, most of the energy being in the kinetic
energy of the associated protons, rather than in photons. 

The most likely way to achieve a nonthermal spectrum in an energetically 
efficient manner is if the kinetic energy of the flow is re-converted into 
random energy
via shocks, after the flow has become optically thin. This is a plausible
scenario, in which two cases can be distinguished. In the first case (a)
the expanding fireball runs into an external medium (the ISM, or a pre-ejected
stellar wind (\cite{rm92,mr93,ka94,sapi95}). The second
possibility (b) is that (even before such external shocks occur) internal
shocks develop in the relativistic wind itself, faster portions of the
flow catching up with the slower portions (\cite{rm94,px94}). In both of its
versions, the fireball shock scenario is completely generic, independent of 
the specific nature of the progenitor, as long as it delivers the appropriate 
amount of energy ($\simg 10^{52}$ erg) in a small enough region ($\siml 10^7$ cm).
This model has been successful in explaining the major observational
properties of the gamma-ray emission, and is the main paradigm used for
interpreting the GRB observations.

External shocks will occur in an impulsive outflow of total energy
$E_o$ in an external medium of average particle density $n_o$ at a radius
\beq
r_{dec} \sim 10^{17}n_o^{1/3} E_{53}^{1/3} \eta_2^{-2/3} ~{\rm cm}~,
\enq
where the lab-frame energy of the swept-up external matter ($\Gamma^2 m_p c^2$
per proton) equals the initial energy $E_o$ of the fireball, and $\eta=\Gamma =
10^2\eta_2$ is the final bulk Lorentz factor of the ejecta.
The typical observer-frame dynamic time at the shock (assuming the cooling
time is shorter than this) is $t_{dec} \sim r_{dec}/c \Gamma^2 \sim$ seconds,
for typical parameters, and the observed burst duration would be $\sim t_{dec}$.
The impulsive assumption requires that the initial energy input occur in a time
shorter than $t_{dec}$. Variability on timescales shorter than $t_{dec}$
may occur on the cooling timescale or on the dynamic timescale for
inhomogeneities in the external medium, but generally external shocks are 
not ideal for reproducing highly variable profiles (\cite{sapi98}).
However, they can reproduce bursts with several peaks (\cite{pm98a})
and may therefore be applicable to the class of long, smooth bursts, or bursts
with a few smooth peaks.
 
In a wind scenario, the energy input occurs non-impulsively over a timescale 
$t_w$, and this can lead to internal shocks at smaller radii where 
$t_{dec} < t_w$.
One expects a similar initial dynamic behavior as in the impulsive injection
case, assuming that a lab-frame luminosity $L_o$ and mass outflow $\dot M_o$ are 
injected at $r\sim r_l$ and are continuously maintained over a time $t_w$.
Initially $\Gamma \propto r$ (with comoving temperature $T \propto r^{-1}$), 
followed by saturation at the value $\Gamma_{max} \sim \eta$ which occurs at 
the radius $r_{sat} \sim /r_l \eta$, where $\eta=L_o/ {\dot M_o c^2}$. In 
such wind models, internal shocks will occur at a radius (\cite{rm94})
\beq
r_{dis} \sim c t_{var} \eta^2 \sim 3\times 10^{14} t_{var} \eta_2^2 ~{\rm cm},
\enq
where shells of different energies $\Delta \eta \sim \eta$ initially separated
by $c t_{var}$ catch up with each other (with $t_{var} < t_w$).
In order for internal shock radius $r_{dis}$  to occur above the wind 
photosphere radius
\beq
r_{ph} \sim {\dot M} \sigma_T /(4\pi m_p c \Gamma^2) 
 \sim 10^{14} L_{53}\eta_2^{-3} \hbox{ cm}, 
\enq
but also at radii greater than the saturation radius $r_{sat} \sim r_l \eta$
(so that most of the energy does not come out in a photospheric 
quasi-thermal radiation component) one needs to have
$7.5\times 10^1 L_{53}^{1/5} t_{var}^{-1/5} \siml \eta \siml 
3\times 10^2 L_{53}^{1/4} t_{var}^{-1/4}$, where $L_{53}=E_{53}/t_w$.
This type of models have the advantage (\cite{rm94}) that they allow an 
arbitrarily complicated light curve, 
the shortest variation timescale $t_{var} \simg 10^{-3}$ s being limited only 
by the dynamic timescale at $r_l$, where the energy input may be expected to
vary chaotically. Such internal shocks have been shown explicitly  to
reproduce (and indeed even be required by) some of the more complicated 
light curves (\cite{sapi98,pm98d}; see however \cite{dermit98}). 

\section{Progenitors and Central Engines}

Even before the measurement of a high redshift in a GRB afterglow, the
difficulty in detecting the host galaxies of bright bursts(e.g.\cite{sch97})
motivated the exploration of ways of increasing the possible total energy
budget of GRB. The first explicit model to do this (\cite{mr97b}), which was
formulated before afterglows and their redshifts were discovered, involved
converting a large fraction of the binding energy of a black hole and torus 
system ($\sim 10^{54}$ ergs) into a fireball through MHD torques
which power a Poynting jet outflow. Such a system would naturally arise
from a \nsns or a \bhns merger (and it may also arise naturally in a failed 
SN Ib or a ``hypernova").
In the last year, a number of other possible energy sources have been 
considered as possible candidates for powering GRB (\cite{fw98,pop98}).
A fact which is not widely realized is that {\it all}
plausible GRB progenitors suggested so far (e.g. NS-NS or NS-BH mergers,
Helium core - black hole [He/BH] or white dwarf - black hole [WD-BH] mergers,
and the wide category labeled as hypernova or collapsars including failed
supernova Ib [SNe Ib], single or binary Wolf-Rayet [WR] collapse, etc.) are
expected to lead to a BH plus debris torus system, and they are all
capable of producing relativistic outflows through the same mechanisms.  
An important point is that the {\it overall energetics} from these various 
progenitors {\it do not differ} by more than about one order of magnitude
(\cite{mrw98b}).
 
Two large reservoirs of energy are available in a generic merger or collapse
scenario: the binding energy of the orbiting debris, and the spin energy of 
the black hole (\cite{mr97b,pac98}).

The first mechanism, relying on the binding energy of the torus, can provide 
up to 42\% of the rest mass energy of the disk, for a maximally rotating 
black hole.  The $\nu\bar\nu \to e^+ e^-$ process can tap the thermal
energy of the torus produced by viscous dissipation. For this mechanism to
be efficient, the neutrinos must escape before being advected into the hole;
on the other hand, the efficiency of conversion into pairs (which scales with
the square of the neutrino density) is low if the neutrino production is too
gradual. Typical estimates suggest a  neutrino powered fireball of 
$\siml 10^{51}$ erg (\cite{ruf97,pop98}), except perhaps in the ``collapsar" 
or failed SN Ib case where estimates (\cite{pop98}) indicate up to 
$10^{52.3}$ ergs for optimum parameters. If the fireball is collimated
into a solid angle $\Omj$ then of course the apparent ``isotropized" energy
would be larger by a factor $(4\pi/\Omj)$ , but unless $\Omj$ is $\siml 10^{-2}
-10^{-3}$ this may fail to satisfy the apparent isotropized energy of
$10^{53.5}$ ergs implied by a redshift $z=3.4$ for GRB 971214 (a similar
energy budget is implied by a possible photometric redshift $z\sim 5$ for
GRB 980329, \cite{fru98}).
An alternative way to tap the torus energy is through dissipation of magnetic
fields generated by the differential rotation in the torus 
(\cite{pac91,nar92,mr97b,ka97}). 
Even before the BH forms, a NS-NS merging system might lead to winding up of 
the fields and dissipation in the last stages before the merger
(\cite{mr92,vie97a}).

The second mechanism relies on tapping the spin energy of the black hole itself,
and it can in principle extract up to 29\% of the rest mass energy of the 
black hole. A hole formed from a coalescing compact binary is guaranteed to be
rapidly spinning, and, being more massive, could contain more energy than
the torus; the energy extractable in principle through MHD coupling to the
rotation of the hole by the \cite{bz77} effect could then be even larger than 
that contained in the orbiting debris (\cite{mr97b,pac98}). Collectively, any 
such MHD outflows have been referred to as Poynting jets.

The various progenitors (\nsns, \bhns, failed SNe, hypernovae and
various collapsars) differ only slightly in the 
mass of the BH and that of the debris torus they produce, and they may differ 
more markedly in the amount of rotational energy contained in the BH. Strong 
magnetic fields, of order $10^{15}$ G, are needed needed to carry away the
rotational or gravitational energy in a time scale of tens of seconds
(\cite{us94,tho94}).  

If the magnetic fields do not thread the BH,
then a Poynting outflow can at most carry the gravitational binding energy
of the torus. For a maximally rotating and for a non-rotating BH this is
0.42 and 0.06 of the torus rest mass, respectively. The torus or disk mass
in a NS-NS merger is (\cite{ruf97}) $M_d\sim 0.1\msun$ , and for a
NS-BH, a He-BH, WD-BH merger or a binary WR collapse it may be estimated at
(\cite{pac98,fw98}) $M_d \sim 1\msun$.  In the HeWD-BH merger
and WR collapse the mass of the disk is uncertain due to lack of
calculations on continued accretion from the envelope, so $1\msun$ is just
a rough estimate. The energy available is then
\beq
E_{max,t} \sim 
\cases{ 8\times 10^{53} \eps (M_d/\msun)~\hbox{ergs}~,& (fast rot.);\cr
         1.2\times 10^{53}\eps (M_d/\msun)~\hbox{ergs}~,& (slow rot.),\cr} .
\label{eq:edisk}
\enq
where $\eps$ is the efficiency for converting gravitational
into MHD jet energy.  The largest energy reservoir is therefore likely
to be associated with NS-BH, HeWD-BH or binary WR collapse, which have
larger disks and fast rotation, the maximum energy being $\sim 8 \times
10^{53} \eps (M_d/\msun)$ ergs; for the (fast rotating but smaller disk) 
NS-NS merger it is $\sim 8\times 10^{52} \eps (M_d/0.1 \msun) $ ergs;
and for the failed SNe Ib (which is a slow rotator) it is $\sim 1.2\times 
10^{53}\eps (M_d/\msun)$ ergs, Conditions for the efficient escape of a
high-$\Gamma$ jet may, however, be less propitious if the ``engine" is
surrounded by an extensive envelope, which is the case in the failed SNE Ib
or hypernova models.
 
If the magnetic fields in the torus thread the BH, the spin energy
of the BH can be extracted via the \cite{bz77} (B-Z) mechanism (\cite{mr97b}). 
The extractable energy is 
\beq
E \sim \eps f(a)\Mbh c^2~,
\enq
where $\eps$ is the MHD efficiency factor and $a = Jc/G M^2$ is the
rotation parameter, which equals 1 for a maximally rotating black hole.
The rotational parameter $f(a)=1-([1+\sqrt{1-a^2}]/2 )^{1/2}$ 
is small unless $a$
is close to 1, where it sharply rises to its maximum value $f(1)=0.29$,
so the main requirement is a rapidly rotating black hole, $a \simg 0.5$.
For a maximally rotating BH, the extractable energy is therefore
\beq
E_{max,bh}\sim 
0.29 \eps\Mbh c^2 \sim 5\times 10^{53}\eps (\Mbh/\msun)~\hbox{ergs}.
\enq
Rapid rotation is essentially guaranteed in a  NS-NS merger, since
the radius (especially for a soft equation of state) is close to that of a
black hole and the final orbital spin period is close to the required
maximal spin rotation period. Since the central BH will have a mass 
(\cite{ruf97,ruja98}) of about $2.5 \msun$, the NS-NS system can thus power
a jet of up to 
\beq
E_{max,bh} \sim 1.3 \times 10^{54} \eps (\Mbh/2.5\msun)~\hbox{ergs}~.
\label{nsbz}
\enq
A maximal rotation rate may also be possible in a He-BH merger, depending
on what fraction of the He core gets accreted along the rotation axis as
opposed to along the equator (\cite{fw98}), and the same should
apply to the binary fast-rotating WR scenario, which probably does not
differ much in its final details from the He-BH merger. For a fast rotating 
BH of $2.5-3\msun$ threaded by the magnetic field, the maximal energy
carried out by the jet is then similar or somewhat larger than in 
equation (\ref{nsbz}).
The scenarios less likely to produce a fast rotating BH are the NS-BH merger
(where the rotation parameter could be limited to $a \leq M_{ns}/\Mbh$,
unless the BH is already fast-rotating) and the failed SNe Ib (where the
last material to fall in would have maximum angular momentum, but the
material that was initially close to the hole has less angular momentum).
The electromagnetic energy extraction from the BH in these could be limited
by the $f(a)$ factor, but a lower limit would be given by the energy
available from the gravitational energy of the disk, in the second line
of equation (\ref{eq:edisk}).

Thus in the case of a Poynting (MHD) jet powered by the binding energy of 
the torus, the total energetics between the various models differs at most 
by a factor 20, whereas for Poynting jets powered by the spin energy of the
black hole they differ by at most a factor of a few, depending on the
rotation parameter. For instance, even allowing for low total efficiency
(say 30\%), a NS-NS merger whose jet is powered by the torus binding energy
would only require a modest beaming of the $\gamma$-rays by a factor
$(4\pi/\Omj)\sim 20$, or no beaming if the jet is powered by the B-Z mechanism,
to produce the equivalent of an isotropic energy of $10^{53.5}$ ergs.  The
beaming requirements of BH-NS and some of the other progenitor scenarios are
even less constraining. Thus, even extreme redshifts $z\sim 5$ such as inferred
by \cite{fru98} can be easily satisfied by secanrios leading to a BH plus torus
system.

An interesting case is the apparent coincidence of GRB 980425 with the SN Ib/Ic 
1998bw (\cite{gal98}).
A simple but radical interpretation (\cite{wawe98}) 
is that all GRB may be
associated with SNe Ib/Ic and differences arise only from different
viewing angles relative to a very narrow jet. The difficulties with this
are that it would require extreme collimations by factors
$10^{-3}-10^{-4}$, and that the statistical association of any subgroup of
GRB with SNe Ib/Ic (or any other class of objects, for that matter) is so
far not significant (\cite{kip98}).
If however the GRB 980425/1998bw association is real (\cite{wes98}),
then we may be in the presence of a new subclass of GRB with lower energy
$E_\gamma \sim 10^{48} (\Omj /4\pi )$ erg, which is only rarely observable
even though its comoving volume density could be substantial. In this,
more likely interpretation, the great majority of the observed GRB would
have the energies $E_\gamma \sim 10^{54}(\Omj/4\pi)$ ergs as inferred from
high redshift observations.
Nonetheless, until further examples of such associations are discovered, one 
may not entirely discount the possibility of this being an extremely rare,
low probability chance coincidence.

\section{Afterglows: the Simple ``Standard" Model}

One can understand the dynamics of the afterglows of gamma-ray bursts in
a fairly simple manner, without worrying about the uncertainties of the 
progenitor. This can be done through a relativistic generalization 
of the method used to understand supernova remnants without fully understanding 
the initiating explosion.  The simplest hypothesis is that the afterglow is due 
to a relativistic expanding blast wave, which decelerates as time goes on
(\cite{mr97a}; earlier simplified discussions were given by 
\cite{ka94b,pacro93,rm92}).  The
complex time structure of some bursts suggests that the central
trigger may continue for up to 100 seconds. However, at much later
times all memory of the initial time structure would be lost:
essentially all that matters is how much energy and momentum has been
injected; the injection can be regarded as instantaneous in the
context of the much longer afterglow. Detailed calculations and predictions 
from such a model (\cite{mr97a}) preceded the observations of the first 
afterglow detected, GRB970228 (\cite{cos97,jvp97}).
 
The simplest spherical afterglow model consist of a three-pieced power law
spectrum with two breaks. At low frequencies there is a steeply rising 
synchrotron self-absorbed spectrum up to a self-absorption break $\nu_a$, 
followed by a +1/3 energy index spectrum up to the synchrotron break $\nu_m$
corresponding to the minimum energy $\gamma_m$ of the power-law accelerated 
electrons, and then a $-(p-1)/2$ energy spectrum above this break, 
for electrons in the adiabatic regime (where $\gamma^{-p}$ is the electron 
energy distribution above $\gamma_m$). In addition, a third break is expected 
at energies where the electron cooling time becomes short compared to the 
expansion time, with a spectral slope $-p/2$ above that. With the inclusion 
of this third, ``cooling" break $\nu_b$, first calculated in \cite{mrw98} and 
more explicitly detailed in \cite{spn98}, 
one has what has come to be called the simple ``standard" model of a GRB 
afterglow. This implicitly assumes spherical symmetry (for a jet with 
opening angle $\theta_j$ this remains valid as long as $\Gamma \simg 
\theta_j^{-1}$) and an impulsive energy input. As the remnant expands the 
spectrum simply moves to lower frequencies, so that in a given band the flux 
decays as a power law in time, whose index can change as breaks move through 
the observing band.

The simple standard model has been remarkably successful at
explaining the gross features of the GRB 970228, GRB 970508 and other
afterglows (\cite{wrm97,tav97,wax97,rei97}).
The multi-wavelength data analysis has in fact advanced to the point where one
can use observed light curves at different times to extrapolate in time to get
spectral snapshots at a fixed time (\cite{wax97,wiga98}), allowing fits for 
the different physical 
parameters of the burst and environment, such as the total energy $E$, the 
magnetic and electron-proton coupling parameters ${\eps}_B$ and ${\eps}_e$ and 
the external density $n_o$. This has led to the temptation to take the assumed 
sphericity for granted. For instance, the lack of a break in the late light 
curve of GRB 970508 prompted the inference (\cite{ram98}) that all afterglows are 
essentially isotropic, leading to the very large (isotropic) energy estimate 
of $10^{53.5}$ ergs in GRB 971214. However, what these fits constrain is only 
the energy per unit solid angle ${\cal E}= (E/\Omj)$. Furthermore,
the assumption that a break should have been seen after months due to the
fireball having become nonrelativistic, or $\Gamma$ having dropped below an 
inverse jet opening angle size, is based upon the simple impusive energy 
input (a delta, or top hat function in initial energy and $\Gamma$, which 
are independent of angle). These are useful simplifications, but it is easy
to see that departures from it are natural and would certainly not be 
surprising. Thus, as we emphasize below, there are so far {\it no strong 
constraints} on the possible anisotropy of the outflow at various energies.

\section{Realistic Afterglows Models}

In the simplest departure from a spherical model the blast wave energy may 
be channeled into a solid angle $\Omj$. In this case one expects 
(Rhoads, 1997a, 1997b)
a faster decay of $\Gamma$ after sideways 
expansion sets in, and a decrease in the brightness is expected after the edges
of the jet become visible, when $\Gamma$ drops below $\Omj^{-1/2}$. 
A simple calculation using the usual scaling laws leads indeed to a steepening 
of the flux power law in time.  The lack of such an observed downturn in 
the optical can, in a first approximation, be interpreted as an indication
of the sphericity of the late stages of the fireball, and by extrapolation to
the entire fireball including its early, gamma-ray emitting stages. 

There are, however, several important caveats. The first one is
that the above argument assumes a simple, impulsive energy input (lasting
$\siml$ than the observed $\gamma$-ray pulse duration), characterized by a
single energy and bulk Lorentz factor value. Estimates for the time needed
to reach the non-relativistic regime, or $\Gamma < \Omega_j^{-1/2} \siml$
few, could then be under a month (\cite{vie97a}),
especially if an initial radiative regime with $\Gamma\propto r^{-3}$
prevails. It is unclear whether, even when electron radiative time scales
are shorter than the expansion time, such a regime applies, as it would
require strong electron-proton coupling (\cite{mrw98}).

Furthermore, even the simplest reasonable departures from a
top-hat approximation (e.g. having more energy emitted with lower Lorentz
factors at later times, which still do not exceed the gamma-ray pulse
duration) would drastically extend the afterglow lifetime in the
relativistic regime, by providing a late ``energy refreshment" to the
blast wave on time scales comparable to the afterglow time scale (\cite{rm98}).
The transition to the $\Gamma < \Omega_j^{-1/2}$ regime
occurring at $\Gamma\sim$ few could then occur as late as six months to
more than a year after the outburst, depending on details of the brief
energy input.  Even in a simple top-hat model, more detailed calculations
show that the transition to the non-relativistic regime is very gradual
($\delta t/t \simg 2$) in the light curve. A numercial computation of the
sideways expansion effects also shows that its effects are not so drastic as 
inferred from simple scaling for the material along the line of sight. This is
because even though the flux from the head-on part of the remnant decreases 
faster, this is more than compensated by the increased emission measure from 
sweeping up external matter over a larger angle, and by the fact that the 
extra radiation,
which arises at larger angles, arrives later and re-fills the steeper
light curve. The sideways expansion thus actually can slow down the flux
decay (\cite{pm98c}) rather than making for a faster decay.
Thus the conclusion is that we {\it do not} yet have significant evidence 
for whether the outflow is jet-like. The lack of a noticeable downturn in
the light-curve is therefore, so far, compatible with either a spherical 
or a jet-like outflow.

The ratio $L_\gamma/L_{opt}$ (or $L_\gamma / L_x$) can be quite different 
from burst to burst. The fit of \cite{wiga98} for GRB 970508 
indicates an afterglow (X-ray energies or softer)
energy per solid angle ${\cal E}_{52} =3.7$, while at $z=0.835$ with $h_{70}=1$
the corresponding $\gamma$-ray ${\cal E}_{52\gamma} =0.63$. On the other hand
for GRB 971214, at $z=3.4$, the numbers are ${\cal E}_{52} = 0.68$ and
${\cal E}_{52\gamma}=20$.  The $\gamma$-ray bursts themselves require ejecta 
with $\Gamma > 100$. The gamma-rays we receive come only from material whose 
motion is directed within one degree of our line of sight.  They therefore 
provide no information about the ejecta in other directions: the outflow 
could be isotropic, or concentrated in a cone of angle (say) 20 degrees
(provided that the line of sight lay inside the cone).  At observer
times of more than a week, the blast wave would be decelerated to a
moderate Lorentz factor, irrespective of the initial value. The
beaming and aberration effects are less extreme so we observe
afterglow emission not just from material moving almost directly
towards us, but from a wider range of angles.
 
The afterglow is thus a probe for the geometry of the ejecta ---
at late stages, if the outflow is beamed, we expect a spherically-symmetric
assumption to be inadequate; the deviations from the predictions of such a
model would then tell us about the ejection in directions away from our line
of sight.  It is quite possible, for instance, that there is relativistic
outflow with lower $\Gamma$ (heavier loading of baryons) in other directions;
this slower matter could even carry most of the energy (\cite{wrm97,pac98}). 
This hypothesis is, in fact, supported to some degree by the fits of 
\cite{wiga98} mentioned above.

\section{Environment Effects, Evolution and Diversity}

One expects afterglows to show a significant amount of diversity. This
is expected both because of a possible spread in the total energies (or 
energies per solid angle as seen by a given observer), a possible spread 
or changes in the injected bulk Lorenzt factors, and also from the 
fact that GRB may be going off in very different environments.

The angular dependence of the outflow, and the radial dependence of the
density of the external environment can have a marked effect on the time
dependence of the observable afterglow quantities (\cite{mrw98}).
So do any changes of the bulk Lorentz factor and energy output during even
a brief energy release episode (\cite{rm98}). The afterglow light curves are 
also affected by the degree of coupling between electrons and protons in
the outflow (\cite{mrw98,pm98b}). Diversity in the light curves of objects 
such as GRB 970508, e.g. sharp rises or humps followed by a renewed decay 
(\cite{ped98,pir98a}) is symptomatic of {\it departures} from the simple 
standard model.  Detailed time-dependent model fits (\cite{pmr98}) to the 
X-ray, optical and radio light curves of GRB 970228 and GRB 970508 show that,
in order to explain the humps, a {\it non-uniform} injection or an 
{\it anisotropic} outflow is required. These fits indicate that 
the shock physics may be a function of the shock strength, and also indicate
that dust absorption may be needed to simultaneously fit the X-ray and 
optical fluxes (the latter being affected more severely).
The effects of beaming (outflow within a
limited range of solid angles) can be significant (\cite{pm98c}), but are
coupled with other effects, and a careful analysis is needed to
disentangle them. 

Spectral signatures, such as atomic edges and lines, may be expected
both from the outflowing ejecta (\cite{mr98a}) and from the external 
medium (\cite{pl98,mr98b}) in the X-ray and optical spectrum of
afterglows. These may be used as diagnostics for  the outflow Lorentz
factor, or as alternative measures of the GRB redshift.
An interesting prediction (\cite{mr98b}; see also \cite{ghi98,bot98}) is that 
the presence of a measurable Fe  K-$\alpha$ {\it emission} line could be a 
diagnostic of a hypernova, since in this case one can expect a massive envelope 
at a radius comparable to a light-day where $\tau_T \siml 1$, capable of 
reprocessing the X-ray continuum by recombination and fluorescence.

The location of the afterglow relative to the host galaxy center can
provide clues both for the nature of the progenitor and for the external 
density encountered by the fireball. A hypernova model would be expected
to occur inside a galaxy, in fact inside a high density ($n_o > 10^3-10^5$).
Some bursts are definitely inside the projected image of the host galaxy, and 
some also show evidence for a dense medium at least in front of the 
afterglow (\cite{ow98}).  On the other hand, for a number of bursts there are 
strong constraints from the lack of a detectable, even faint, host 
galaxy (\cite{sch98}).
In NS-NS mergers one would expect a BH plus debris torus system and 
roughly the same total energy as in a hypernova model, but the mean distance 
traveled from birth is of order several Kpc (\cite{bsp98}),
leading to a burst presumably in a less dense environment. The fits of 
\cite{wiga98} to the observational data on GRB 970508 and 
GRB 971214 in fact suggest external densities in the range of $n_o=$
0.04--0.4 cm$^{-1}$, which would be more typical of a tenuous interstellar 
medium (however, \cite{reila98} report a fit for GRB 980329 with $n_o\sim 10^4$
cm$^{-3}$).  These could arise within the volume of the galaxy, but on
average one would expect as many GRB inside as outside. This is based on
an estimate mean NS-NS merger time of $10^8$ years; other estimated merger
times (e.g. $10^7$ years, \cite{vdh92}) would give a burst much closer to
the birth site. BH-NS mergers would also occur in timescales $\siml 10^7$ 
years, and would be expected to give bursts well inside the host 
galaxy (\cite{bsp98}).

\section{ Conclusions }
 
The simple blast wave model has proved quite robust in providing a 
consistent overall interpretation of the major features of gamma-ray bursts
and their afterglows at various frequencies. However, the constraints on the 
angle-integrated energy, especially in $\gamma$-rays, are not strong, and 
beaming effects remain uncertain. Some of the observed light curve humps
are likely to be indicative of either non-uniform injection episodes or 
anisotropic outflows. A relatively brief (1-100 s), probably modulated 
energy input appears the likeliest interpretation for most bursts,
although in some progenitor scenarios there may be delayed effects.  
This can provide an explanation both for highly variable $\gamma$-ray light 
curves and late glitches in the afterglow decays.
One needs to be mindful of the possibility of there being more subclasses 
of classical GRB than just short ones and long ones. For instance, GRB with 
no high energy pulses (NHE) appear to have a different (but still isotropic) 
spatial distribution than those with high energy (HE) pulses (\cite{pen96}).
Some caution is needed in interpreting this, since selection effects could 
lead to a bias against detecting HE emission in dim bursts (\cite{nor98}).
A connection to peculiar supernova events, or the formation of a temporarily 
rotationally stabilized strong-field pulsar are also possibilities which 
cannot be ruled out,

Significant progress has been made in understanding how gamma-rays can arise in
fireballs produced by brief events depositing a large amount of energy in a
small volume, and in deriving the generic properties of the long wavelength
afterglows that follow from this.  There still remain a number of mysteries,
especially concerning the identity of their progenitors, the nature of the 
triggering mechanism, the transport of
the energy and the time scales involved. Nevertheless,
even if we do not yet understand the intrinsic gamma-ray burst central engine,
they may be the most powerful beacons for probing the high redshift ($z > 5$)
universe. Even if their total energy is reduced by beaming to a ``modest"
$\sim 10^{52}-10^{52.5}$ ergs in photons, they are the most extreme
phenomena that we know about in high energy astrophysics.  The modeling
of the burst mechanism itself will continue to be a major challenge to
theorists and to computational techniques. However, there is every prospect
for continued and vigorous developments both in the observational analyses
and in the theoretical understanding of these fascinating objects.
 
\acknowledgements
{I am grateful to Martin Rees for stimulating collaborations
on this subject, as well as to Ralph Wijers, Hara
Papathanassiou and Alin Panaitescu. This research is supported in
part by NASA NAG5-2857}


\vskip 0.4cm

\begin{thebibliography}{}
\small
\bibitem[Blandford \& Znajek, 1977]{bz77} Blandford, R.D. \& Znajek, R.L., 1977, MNRAS, 179, 433
\bibitem[Bloom, Sigurdsson \& Pols, 1998]{bsp98} Bloom, J., Sigurdsson, S. \& Pols, O., 1998, MNRAS in press (astro-ph/9805222)
\bibitem[B\"ottcher, \etal, 1998]{bot98} B\"ottcher, M, \etal, 1998, astro-ph/9809156
\bibitem[Costa et al, 1997]{cos97} Costa, E., et al., 1997, Nature, 387, 783 
\bibitem[Dermer \& Mitman, 1998]{dermit98} Dermer, C \& Mitman, K, 1998, astro-ph/9809411
\bibitem[Djorgovski, \etal, 1998]{djo98} Djorgovski, S.G. \etal, 1998, astro-ph/9808188 
\bibitem[Eichler, \etal, 1989]{eic89} Eichler, D., Livio, M., Piran, T. and Schramm, D.N., 1989, Nature, 340, 126
\bibitem[Fishman \& Meegan, 1998]{fm95} Fishman, G. \& Meegan, C., 1995, Ann.Rev Astr.Ap.,33, 415
\bibitem[Fruchter, 1998]{fru98} Fruchter, A, asstro-ph/9810224
\bibitem[Fryer \& Woosley, 1998]{fw98} Fryer, C \& Woosley, S, 1998, ApJ(Lett) subm (astro-ph/9804167
\bibitem[Galama, \etal, 1998]{gal98} Galama, T. et al., 1998, Nature, in press (astro-ph/9806175)
\bibitem[Ghisellini, \etal, 1998]{ghi98} Ghisellini, G, \etal, 198, astro-ph/9808156
\bibitem[Goodman, 1986]{goo86} Goodman, J., 1986, Ap.J.(Lett.), 1986, 308, L47
\bibitem[Groot et al, 1997]{gro97} Groot, P. et al., 1997;  in {\it Gamma-Ray Bursts}, Meegan, C., Preece, R. \& Koshut, T., eds., 1997 (AIP: New York), p. 557
\bibitem[Harding \& Baring, 1994]{hb94} Harding, A.K. and Baring, M.G., 1994, in {\it Gamma-ray Bursts},  ed.  G. Fishman, \etal, p. 520 (AIP 307, NY)
\bibitem[Horack, \etal, 1996]{hor96} Horack, J. M., Mallozzi, R. S., \& Korshut, T. M. 1996, ApJ, 466, 21
\bibitem[Katz, 1994]{ka94} Katz, J., 1994, ApJ, 422, 248
\bibitem[Katz, 1994b]{ka94b} Katz, J., 1994b, ApJ, 432, L107
\bibitem[Katz, 1997]{ka97} Katz, J.I., 1997, ApJ, 490, 633
\bibitem[Kippen, \etal, 1998]{kip98} Kippen, R.M. et al., 1998, ApJ subm (astro-ph/9806364)
\bibitem[Krumholz, \etal, 1998]{kru98} Krumholz, M, Thorsett, S \& Harrison, F, 1998, preprint (astro-ph/9807117)
\bibitem[Kulkarni, et al, 1998]{kul98} Kulkarni, S., et al., 1998, Nature, 393, 35
\bibitem[\Mesz, 1995]{m95} \Mesz, P., 1995, 17th Texas Symp. Relativistic Astrophysics, H. B\"ohringer et al, N.Y. Acad. Sci., 440
\bibitem[\Mesz \& Rees, 1992]{mr92} \Mesz, P \& Rees, M.J., 1992, ApJ, 397, 570
\bibitem[\Mesz \& Rees, 1992b]{mr92b} \Mesz, P \& Rees, MJ, 1992b, MNRAS, 257, 29P
\bibitem[\Mesz \& Rees, 1993]{mr93} \Mesz, P. \& Rees, M.J., 1993, ApJ, 405, 278
\bibitem[\Mesz \& Rees, 1997a]{mr97a} \Mesz, P \& Rees, M.J., 1997a, ApJ, 476, 232
\bibitem[\Mesz \& Rees, 1997b]{mr97b} \Mesz, P \& Rees, M.J., 1997b, ApJ, 482, L29
\bibitem[\Mesz \& Rees, 1998a]{mr98a} \Mesz, P \& Rees, M.J., 1998a, ApJ(Letters) 502, L105
\bibitem[\Mesz \& Rees, 1998b]{mr98b} \Mesz, P \& Rees, M.J., 1998b, MNRAS, 299, L10  (astro-ph/9806183)
\bibitem[\Mesz, Rees \& Wijers, 1998]{mrw98} \Mesz, P, Rees, M.J. \& Wijers, R, 1998, Ap.J., 499, 301 (astro-ph/9709273)
\bibitem[\Mesz, Rees \& Wijers, 1998b]{mrw98b} \Mesz, P, Rees, M.J. \& Wijers, R, 1998b, New Astron, in press (astro-ph/9808106)
\bibitem[Metzger et al, 1997]{metz97} Metzger, M et al., 1997, Nature, 387, 878
\bibitem[Murakami, et al, 1997]{mur97} Murakami, T. et al., 1997, in {Gamma-Ray Bursts}, Meegan, C., Preece, R. \& Koshut, T., eds., 1997 (AIP: New York), p. 435
\bibitem[Narayan, Piran \& Shemi, 1991]{nar91} Narayan, R., Piran, T. \& Shemi, A, 1991, Ap.J.(Lett.), 379, L17
\bibitem[Narayan, \Pacz \& Piran, 1992]{nar92}Narayan, R., \Pacz, B. \& Piran, T., 1992, Ap.J., 395, L83
\bibitem[Norris, 1998]{nor98} Norris, J, 1998, private communication
\bibitem[Norris, \etal, 1995]{nor95} Norris, J. P., et al. 1995, ApJ, 439, 542
\bibitem[Owen, \etal, 1998]{ow98} Owen, A., et al, 1998, Astron.\&Astrophys. in press (astro-ph/9809356),
\bibitem[\Pacz \& Rhoads, 1993]{pacro93} \Pacz, B. \& Rhoads, J, 1993, Ap.J., 418, L5
\bibitem[\Pacz \& Xu, 1994]{px94} \Pacz, B. \& Xu, G., 1994, ApJ, 427, 708
\bibitem[\Pacz, 1986]{pac86} \Pacz, B., 1986, Ap.J.(Lett.), 308, L43
\bibitem[\Pacz, 1992]{pac91} Paczi\'nski, B., 1991, Acta. Astron., 41, 257
\bibitem[\Pacz, 1998]{pac98} \Pacz, B., 1998, ApJ, 494, L45
\bibitem[Panaitescu \& \Mesz, 1998a]{pm98a} Panaitescu, A \& \Mesz, P, 1998a, ApJ, 492, 683
\bibitem[Panaitescu \& \Mesz, 1998b]{pm98b} Panaitescu, A. \& \Mesz, P., 1998b, ApJ, 501, 772
\bibitem[Panaitescu \& \Mesz, 1998c]{pm98c} Panaitescu, A. \& \Mesz, P., 1998c, ApJ, subm. (astro-ph/9806016)
\bibitem[Panaitescu \& \Mesz, 1998d]{pm98d} Panaitescu, A. \& \Mesz, P., 1998d, ApJ, subm (astro-ph/9810258)
\bibitem[Panaitescu, \Mesz \& Rees, 1998]{pmr98} Panaitescu, A, \Mesz, P \& Rees, MJ, 1998, ApJ, 503, 314
\bibitem[Pedersen, \etal, 1998]{ped98} ApJ 496 (1998) 311
\bibitem[Pendleton, \etal, 1996]{pen96} Pendleton, G, et al., 1996, Ap[J, 464, 606
\bibitem[Perna \& Loeb, 1998]{pl98} Perna, R. \& Loeb, A., 1998, ApJ(Letters), 503, L135
\bibitem[Phinney, 1991]{phi91} Phinney, E.S., 1991, Ap.J.(Lett.), 380, L17
\bibitem[Piro, \etal, 1998]{pir98a} Piro, L, \etal, 1998, A \& A, 331, L41
\bibitem[Popham, Woosley \& Fryer, 1998]{pop98} Popham, R., Woosley, S \& Fryer, C., 1998, ApJ subm (astro-ph/9807028)
\bibitem[Ramaprakash, A, \etal, 1998]{ram98} Ramprakash, A.N. \etal, 1998, Nature, 393, 38
\bibitem[Rees \& \Mesz, 1992]{rm92} Rees, M.J. \& \Mesz, P., 1992, MNRAS, 258, P41
\bibitem[Rees \& \Mesz, 1994]{rm94} Rees, M.J. \& \Mesz, P., 1994, ApJ, 430, L93
\bibitem[Rees \& \Mesz, 1998]{rm98} Rees, M.J. \& \Mesz, P., 1998, ApJ, Letters, 496, L1
\bibitem[Reichart \& \Mesz, 1997]{rm97} Reichart, D. \& \Mesz, P, 1997, ApJ, 483, 597
\bibitem[Reichart, 1997]{rei97} Reichart, D., 1997, ApJ, in press
\bibitem[Reichart, 1998]{rei98} Reichart, D., 1998, ApJ, 495, L99
\bibitem[Reichart \& Lamb, 1998]{reila98} Reichart, D. \& Lamb, D.Q., 1998, talk at the Rome Conference on GRB in the Afterglow Era.
\bibitem[Rhoads, 1997a]{rho97a} Rhoads, J, 1997a, Ap.J., 487, L1
\bibitem[Rhoads, 1997b]{rho97b} Rhoads, J, 1997b, preprint
\bibitem[Ruffert, \etal, 1997]{ruf97} Ruffert, M., Janka, H.-T., Takahashi, K., Schaefer, G., 1997, A\& A, 319, 122
\bibitem[Ruffert \& Janka, 1998]{ruja98} Ruffert, M. \& Janka, H.-T., 1998, A\&A subm (astro-ph/9804132)
\bibitem[Sari \& Piran, 1995]{sapi95} Sari, R. \& Piran, T., 1995, ApJ, 455, L143
\bibitem[Sari \& Piran, 1998]{sapi98} Sari, R \& Piran, T, 1998, ApJ, 485, 270
\bibitem[Sari, Piran \& Narayan, 1998]{spn98} Sari, R , Piran, T \& Narayan, R, 1998, ApJ, 497, L17 (astro-ph/9712005)
\bibitem[Schaefer, 1997]{sch97} Schaefer, B.E. et al, 1997, ApJ, 489, 693
\bibitem[Schaefer, 1998]{sch98} Schaefer, B.E., 1998, ApJ in press (astro-ph/9810424)
\bibitem[Shemi \& Piran, 1990]{sp90} Shemi, A. and Piran, T., 1990, Ap.J.(Lett.), 365, L55
\bibitem[Tavani, 1997]{tav97} Tavani, M., 1997, ApJ, 483, L87
\bibitem[Taylor, \etal, 1997]{tay97} Taylor, G.B., et al., 1997, Nature, 389, 263
\bibitem[Thompson, 1994]{tho94} Thompson, C., 1994, MNRAS, 270, 480
\bibitem[Totani, 1997]{tot97} Totani, T, 1997, ApJ, 486, L71
\bibitem[Usov, 1994]{us94} Usov, V.V., 1994, MNRAS, 267, 1035
\bibitem[van den Heuvel, 1992]{vdh92} van den Heuvel, E., in {\it X-ray Binaries and Recycled Pulsars} (ed. E.P.J.van den Heuvel and S.A.Rappaport), Kluwer Ac. Pub., 1992, pp 233-256.
\bibitem[van Paradijs, \etal, 1997]{jvp97} van Paradijs, J, \etal, 1997, Nature,386, 686 
\bibitem[Vietri, 1997a]{vie97a} Vietri, M., 1997a, ApJ, 478, L9
\bibitem[Wang \& Wheeler, 1998]{wawe98} Wang, L. \& Wheeler, J.C., 1998, ApJ subm (astro-ph/9806212)
\bibitem[Waxman, 1997]{wax97} Waxman, E., 1997, ApJ(Letters)
\bibitem[Wijers \& Galama, 1998]{wiga98} Wijers, R.A.M.J. \& Galama, T., 1998, ApJ, subm (astro-ph/9805341)
\bibitem[Wijers \etal, 1998]{wij98}  Wijers, R, Bloom, J, Bagla, J \& Natarajan, P, 1998, MNRAS, 294, L17
\bibitem[Wijers, Rees \& \Mesz, 1997]{wrm97} Wijers, R.A.M.J., Rees, M.J. \& \Mesz, P., 1997, MNRAS, 288, L51
\bibitem[Woosley, 1993]{woo93}  Woosley, S., 1993, Ap.J., 405, 273
\bibitem[Woosley, Eastman \& Schmidt, 1998]{wes98} Woosley, S., Eastman, R. \& Schmidt, B., 1998, ApJ, subm (astro-ph/9806299)

\end{thebibliography}
\end{document}